\documentclass[a4paper,twocolumn,11pt,unpublished,noarxiv]{article}

\pdfoutput=1
\usepackage[utf8]{inputenc}
\usepackage[english]{babel}
\usepackage[T1]{fontenc}
\usepackage{amsmath}
\usepackage{dsfont}
\usepackage{braket}
\usepackage{tikz}
\usepackage{lipsum}
\usepackage{amsmath}
\usepackage{amssymb}
\usepackage{fixme}
\usepackage{todonotes}
\usepackage{macros}
\setuptodonotes{inline}
\usepackage{listings}
\definecolor{santieditcolor}{RGB}{255,255,0}
\definecolor{cifueditcolor}{RGB}{60,120,120}
\definecolor{guidoeditcolor}{RGB}{255,87,51}
\definecolor{arieleditcolor}{RGB}{255,0,255}

\begin{document}

\title{Towards exact algorithmic proofs of maximal mutually unbiased bases sets in arbitrary integer dimension}

\author{Santiago Cifuentes\thanks{Universidad de Buenos Aires, Facultad de Ciencias Exactas y Naturales. Departamento de Computación, Buenos Aires, Argentina} \thanks{CONICET-Universidad de Buenos Aires, Instituto de Ciencias de la Computación (ICC), Buenos Aires, Argentina} \and Nicolás Ciancaglini\footnotemark[2] \thanks{Universidad de Buenos Aires, Facultad de Ciencias Exactas y Naturales. Departamento de Física, Buenos Aires, Argentina}  \and Guido Bellomo\footnotemark[2] \footnotemark[3] \and Santiago Figueira\footnotemark[1] \footnotemark[2] \and Ariel Bendersky\footnotemark[3] \thanks{Quantum Researh Center, Technology Innovation Institute, Abu Dhabi, United Arab Emirates}}

\date{}

\maketitle

\begin{abstract}
In this paper, we explore the concept of Mutually Unbiased Bases (MUBs) in discrete quantum systems. It is known that for dimensions $d$ that are powers of prime numbers, there exists a set of up to $d+1$ bases that form an MUB set. However, the maximum number of MUBs in dimensions that are not powers of prime numbers is not known.

To address this issue, we introduce three algorithms based on First-Order Logic that can determine the maximum number of bases in an MUB set without numerical approximation. Our algorithms can prove this result in finite time, although the required time is impractical. Additionally, we present a heuristic approach to solve the semi-decision problem of determining if there are $k$ MUBs in a given dimension $d$.

As a byproduct of our research, we demonstrate that the maximum number of MUBs in any dimension can be achieved with definable complex parameters, computable complex parameters, and other similar fields.
\end{abstract}

\section{Introduction}

The Uncertainty Principle, famously stated by Heisenberg, outlines the fundamental limitations of quantum mechanics. It highlights the trade-off between the accuracy of knowledge about a quantum system's position and momentum.

In the realm of discrete quantum systems, the concept of MUBs extends the notion of complementarity. Bases $\aBase_1, \dots, \mathcal \aBase_n$ are considered mutually unbiased if the inner product between two states, one from each basis, results in a uniform probability distribution if the bases are distinct and a delta function if they are the same. This means that if a state is prepared in basis $k$ and then measured projectively in basis $k'\neq k$, all results are equally likely.

It has been established \cite{Wootters1989} that the maximum number of MUBs in a given dimension $d$ is less or equal than $d+1$, with equality holding true for dimensions that are powers of a prime number. However, for dimensions that are not a power of a prime, the maximum number of MUBs is not known. Despite extensive research, the maximum number of MUBs in the lowest non-prime power dimension of 6 remains uncertain. Currently, it is known that there are three MUBs in this case, but the possibility of a fourth MUB has not been ruled out.

On a different note, First-Order Logic, also referred to as Predicate Logic, is a system of formal languages that utilizes quantified variables to express statements about various structures. An example of a sentence in First-Order Logic is:
\begin{equation}\label{eqn:exampleintro}
(\forall y)(\exists x)\ x^2=y
\end{equation}
This sentence is true for structures such as $\mathbb R_{>0}$ and $\mathbb C$, but false for others such as $\mathbb R$, $\mathbb N$, and $\mathbb Q$. The versatility of First-Order Logic allows it to be applied to a wide range of theories and structures.

Algorithms for determining the truth value of a given First-Order formula exist for some theories, specifically for the theory of the reals, but not in general. 
This makes First-Order Logic a powerful tool when working with these specific theories. In this paper, we will leverage this property to provide algorithms that can prove the existence (or lack thereof) of $k$ MUBs in dimension $d$ within a finite time. Additionally, we will demonstrate the existence of proofs for the number of MUBs in a given dimension and offer insights into the nature of these proofs, the mathematical concepts involved, and the properties of MUBs in fields other than the complex numbers.

This paper is organized as follows. In section \ref{sec:MUB} we present the problem of MUBs and provide an overview of the known results in this area. In section \ref{sec:FOL}, we introduce First-Order Logic and its relevant concepts, including the First-Order theory of the reals, the existential theory of the reals, and elimination of quantifiers. These concepts serve as the foundation for the exact algorithms we present in section \ref{sec:algo}, which can decide the truth of the sentence ``there exists $k$ MUBs in dimension $d$''. Additionally, in section \ref{sec:heur} we discuss a heuristic algorithm that can be used to prove the non-existence of $k$ MUBs in dimension $d$, although it is not guaranteed to halt in every case. Finally, in section \ref{sec:other} we prove that the maximum number of MUBs in fields other than the complex numbers is equivalent to that of the complex numbers.

\section{Mutually unbiased bases}\label{sec:MUB}
% Definiciones y resultados conocidos

Given $k$ orthonormal bases $\aBase_m=\{\ket{\psi^m_i}, i=1,...,d\}$, for $k\in\left\lbrace 1,...,k \right\rbrace$, we say they are mutually unbiased whenever
\begin{equation}\label{eq:mubs}
	\left|\braket{\psi^m_i|\psi^n_j} \right|^2 = \begin{cases}
\frac{1}{d} &\text{if $m\neq n$} \\
\delta_{ij} &\text{otherwise.}
\end{cases}
\end{equation}

A set $\{\aBase_1,\ldots,\aBase_k\}$ is called a set of mutually unbiased bases if each pair of distinct bases is mutually unbiased. Consider a scenario where we perform two measurements, first in base $\aBase_1$, and then in base $\aBase_2$. Regardless of the outcome of the first measurement, each possible outcome for the second measurement is equally likely. In this sense, they are referred to as \textit{unbiased}. In other words, bases $\aBase_1$ and $\aBase_2$ are maximally noncommutative.

While it has been established since 1989~\cite{Wootters1989} that the maximum number of MUBs in $\mathbb{C}^d$ cannot exceed $d+1$, and that a complete set of MUBs exists when $d$ is a prime power, the upper bound on the number of MUBs remains unknown for all other dimensions. Even for the smallest non-prime-power dimension $d=6$, the existence of a complete set of MUBs is still an open problem, and both numerical~\cite{Butterley2007,Brierley2008,Raynal2011} and analytical~\cite{Brierley2009,Brierley2010,Jaming2009,Paterek2009,Bengtsson2007b} evidence indicates that it is unlikely that such a set exists. 

The concept of MUBs has found multiple applications in quantum information problems such as quantum state estimation~\cite{Ivonovic1981,Adamson2010,Yuan2016,Bae2019,Huang2021}, entanglement detection or certification~\cite{Spengler2012,Paul2016,Friis2019,Bae2022}, and cryptography~\cite{Wu2009,Wehner2010,Coles2017}, among others (refer to~\cite{Bengtsson2007,Durt2010} for more comprehensive reviews). Hence, the inquiry into the existence of a maximal set of mutually unbiased bases in dimension $d$ has become a topic of great relevance. Generally, there exist several explicit methods to construct a complete set of MUBs for prime-power dimensions $d = p^n$, including the use of finite fields~\cite{Wootters1989,Klappenecker2004}, the Heisenberg-Weyl group~\cite{Bandyopadhyay2002}, generalized angular momentum operators~\cite{Kibler2006}, and identities from number theory~\cite{Archer2005}. Furthermore, for special cases where $d = 2^n$ and $d = p^2$, it has been demonstrated that such sets can be constructed in a simple and experimentally accessible manner~\cite{Seyfarth2011,Wiesniak2011}.

\section{First-Order logic}\label{sec:FOL}
% Generalidades. Hay que mejorarlo mucho, es una intro con ayuda de chatgpt
\newcommand{\Lang}{{\mathcal L}}
\newcommand{\Lor}{{\mathcal L}_{\rm or}}
\newcommand{\Struct}{{\mathcal M}}
\newcommand{\R}{{\mathbb R}}
\newcommand{\N}{{\mathbb N}}
\newcommand{\C}{{\mathbb C}}

In this section we give a short and informal introduction to First-Order Logic, and the main concepts needed for this work. For a more formal explanation, the reader is referred to classical textbooks on model theory such as \cite{marker2006model}.

First-Order Logic, also known as Predicate Logic, is a formal language to reason on mathematical structures. First-Order Logic contains equality, Boolean connectives --conjunction ($\land$), disjunction ($\lor$), implication ($\to$), negation ($\lnot$)--, variables, and quantification --“for all” ($\forall$) and “exists” ($\exists$)-- over variables. Depending on the kind of mathematical objects under study, one fixes an appropriate language $\Lang$ with symbols of relations, functions and constants. Hence we talk about {\em $\Lang$-formulas} and {\em $\Lang$-structures}, to make explicit that they are constructed over the language $\Lang$. A First-Order $\Lang$-formula $\phi$ has a truth value provided it is evaluated in 1) an $\Lang$-structure $\Struct$, which is a set of elements $X$ (called domain), and an interpretation of all symbols in $\Lang$ as mathematical objects in $X$: an $n$-ary relation symbol of $\Lang$ is interpreted as a subset of $X^n$, an $n$-ary function symbol of $\Lang$ is interpreted as a function $X^n\to X$ and a constant symbol of $\Lang$ is interpreted as an element of $X$, and 2) a {\em valuation} $v$ which maps all free (i.e.\ non-quantified) occurrences of variables in $\phi$ to elements of $X$.
% Further, also a mapping of all free (i.e.\ non-quantified) variables of $\phi$ to elements of $X$ has to be given. 
For an $\Lang$-structure $\Struct$ and valuation $v$, the $\Lang$-formula $\phi$ is {\em true} in $\Struct,v$ if the mathematical property stated by $\phi$ holds in $\Struct$ when mapping free variables of $\phi$ to the domain of $\Struct$ as stated by $v$ and interpreting the relation, function and constant symbols of $\Lang$ as stated by $\Struct$. In this case, we also say that $\Struct,v$ {\em satisfies} $\phi$. If $\phi$ does not have any free variables then the valuation $v$ above plays no role and we simply talk about $\phi$ being true in $\Struct$ or $\Struct$ satisfying $\phi$. An $\Lang$-formula with no free variables is called {\em $\Lang$-sentence}, and represents a property that any structure may or may not satisfy. 
If $T$ is a set of $\Lang$-sentences then we say that $\Struct$ is a {\em model} of $T$ if $\Struct$ satisfies all the $\Lang$-sentences of $T$. 

\paragraph{Theories}
An {\em $\Lang$-theory} $T$ is simply a set of $\Lang$-sentences. The {\em consequences} of $T$ is the set of all $\Lang$-formulas $\phi$ such that for any $\Lang$-structure $\Struct$, if $\Struct$ satisfies $T$ then $\Struct$ satisfies $\phi$. The consequences of $T$ are also called {\em theorems} of $T$. Some $\Lang$-theories may be axiomatized by a finite or a finitely represented set of axioms, meaning that there is a finite or finitely represented $\Lang$-theory $T’$ (whose formulas are called {\em axioms}) such that the consequences of $T$ and $T’$ coincide. An $\Lang$-theory $T$ is {\em complete} if for any $\Lang$-sentence $\phi$,  either $\phi$ or its negation is a theorem of $T$. In that case, any two models $\Struct, \Struct'$ satisfying $T$ will be \textit{elementary equivalent}: for any $\Lang$-sentence $\phi$, $\Struct$  satisfies $\phi$ if and only if $\Struct'$ satisfies $\phi$.

\paragraph{The language of ordered rings}
The language $\Lor$ of ordered rings contains the relational symbol $<$, function binary symbols $+$, $-$ and $\cdot$, and the constant symbols $0$ and $1$. It is important to notice that the syntactic rules for constructing First-Order formulas over this language allows us to write formulas like 

\begin{equation}\label{eqn:example}
(\forall x)\ a\cdot x^2 +b\cdot x+c>0
\end{equation}
 with free variables $a$, $b$ and $c$ and a (bound) variable $x$ universally quantified\footnote{Strictly speaking, $x^2$ is not an allowed term in $\Lor$, but for any $n\in\N$ we use the standard notation of $x^n$ to denote $x\cdot\ldots\cdot x$ ($n$ times).}, but not formulas like $(\exists x)\ \sin(x)=1$ since $\sin$ is not part of $\Lor$. The validity, of course, depends on which structure the formula is evaluated --recall the example of sentence \eqref{eqn:exampleintro}. Observe that First-Order logic only allows to quantify elements, but not other objects such as sets of elements, polynomials, etc.

% (which is false in $\R$, and hence not in RCF) or $\exists x.x\cdot x \cdot x + 1=0$ (which is true in $\R$, and hence in RCF) but prevent us to quantify over sets of reals or over polynomials, like  $\forall P. \exists x.P(x)=0$. These expressions are simply not formulas. Furthermore, the expression $\forall n. (\exists k. n=2\cdot k+1) \to \exists x. x^n=0$ is also syntactically invalid since exponentiation is not a symbol of $\Lang$. Incidentally, notice also that the fact of naming a variable $n$ or $k$ does not make it a natural or integer variable. Hence $\exists k. n=2\cdot k+1$ is true in $\R,v$ for any valuation $v$, and is true in $\N,v$ only when the $v(n)$ is odd. Although exponentiation is not part of $\Lor$, for a metavariable $n\in \N$, we may use $x^n$ as a short for $x\cdot\ldots\cdot x$ ($n$ times).

\subsection{Real Closed Fields}\label{sec:RCF}
The theory of the Real Closed Fields (RCF) is the set of $\Lor$-sentences 
which contains 1) all the axioms for fields, 2) an axiom stating that no square or sum of squares is equal to $-1$, 3) an axiom stating that any $x$ is either $y$ or $-y$, where $y$ is a square, and 4) an axiom stating that any polynomial of odd degree has a root. It is not difficult to see that all these properties can be expressed as $\Lor$-sentences. Alfred Tarski showed that RCF is a complete theory and, since $\R$ is a model of RCF, any $\Lor$-sentence $\phi$ is a consequence of RCF if and only if $\phi$ is true in $\R$. 

\paragraph{Quantifier elimination}
A \textit{decision problem} is a problem that can be formulated as a question with a binary answer \textit{yes} or \textit{no} depending on a (finite) set of input parameters. A decision problem is {\em decidable} if there exists an algorithm which is capable of computing the decision output for each input parameter. 

The problem of deciding whether an $\Lor$-sentence is a consequence of RCF (and, therefore, deciding if it is true over $\R$) is decidable, since it is possible to exhaustively enumerate all theorems of RCF by applying proper derivation rules until either $\phi$ or its negation occurs. However, this algorithm is extremely impractical regarding both the memory and time that it requires. 

Nonetheless, RCF has one more interesting property, also shown by Tarski: {\em quantifier elimination}. This means that for any $\Lor$-formula $\phi$ 
% with free-variables $\bar x=x_1,\dots,x_n$ 
there is a quantifier-free formula $\psi$ 
% with free-variables $\bar x$ 
which is {\em equivalent} to $\phi$ in RCF; in other words, for any valuation $v$, we have that $\phi$ is true in $\R,v$ if and only if $\psi$ is true in $\R,v$. For example, a formula like \eqref{eqn:example} is equivalent in RCF to\footnote{For $n\in\N$ the term $n$ is a short for the $\Lor$-term $1+\dots+1$ ($n$ times)} 
\begin{eqnarray*}
(a=0 \land b=0 \land c>0)& \lor \\
(a\geq0 \land b=0\land c>0 \land 4\cdot a\cdot c>b^2) & \lor \\
(a>0 \land 4\cdot a\cdot c>b^2)&
\end{eqnarray*}
Furthermore, there is an algorithm that transforms any input $\Lor$-formula $\phi$ into a equivalent quantifier-free formula $\psi$. This suffices to determine if $\phi$ is true in $\R$ for a given valuation $v$, because it is easy to evaluate the quantifier-free formula $\psi$ over $v$. 

The importance of quantifier elimination is that it allows for real computer implementations (at least compared with the algorithm based on enumerating all theorems of RCF). The most popular algorithm for quantifier elimination in the theory of Real Closed Fields is the cylindrical algebraic decomposition (CAD) algorithm, which was developed mainly by George E.\ Collins in 1975 \cite{collins1975quantifier}. This algorithm is doubly exponential in the number of variables of the input formula, and is implemented in many popular computer algebra systems, such as Maple, Mathematica or Singular. Also, it is one of the most important algorithms from the field of computational algebraic geometry and has received a lot of attention and improvements since its origin.

Moreover, regarding the computational complexity of these tasks, it is known that when restricted to existential formulas (those of the form $(\exists x_1),\dots,(\exists x_n)\ \psi$, for $\psi$ free of quantifiers), the problem of deciding whether a $\Lor$-formula is true is \textsc{NP-hard} and belongs to \textsc{PSPACE} \cite{papadimitriou2003computational} (namely, it is decidable by a deterministic Turing machine that uses $O(n^k)$ many cells of memory, where $n$ is the size of the input formula and $k$ is fixed). The same applies if the number of blocks of quantifiers of the input formula is fixed \cite{Basu2006}.	

\paragraph{Our setting}
For our developments we will need to express properties on the complex numbers $\C$. Obviously RCF cannot be used straightforwardly since truths in $\R$ do not coincide with truths on $\C$ (e.g.\ axiom 2 of RCF stated above). However, the properties $\phi$ on $\C$ that we will use can be expressed as $\Lor$-formulas by duplicating each complex variable $z$ to match the meaning of the real and imaginary part of $z$. For example, if $z$ is a complex variable, the property $|z|>3$ can be translated to the $\Lor$-sentence $z_1^2+z_2^2>3^2$ by identifying $z$ with a pair of real variables $(z_1,z_2)$.
% ∫\footnote{Strictly speaking, $z_1^2+z_2^2>3^2$ is not in $\Lor$, but can be straightforwardly converted to one of $\Lor$ by the standard convention that $n=1+\dots+1$ ($n$ times) and $w^n=w\cdot \ldots\cdot w$ ($n$ times) for any $n\in\N$.} 
We can translate in this way a number of properties on the complex numbers as properties on the reals. Observe, however, that trigonometric functions cannot be mapped straightforwardly as they are not part of $\Lor$.

\section{Exact algorithms to prove the existence of $k$ MUBs in dimension $d$}\label{sec:algo}
% Dar los tres algoritmos:
% Rcoontra Ultra lento: Enumerar teoremas hasta encontrar la existencia o no de las MUBs.
% Ultra lento: Eliminación de cuantificadores a la Tarski
% Lentísimo: Eliminación de cuantificadores en la teoría existencial de los reales

Proving the existence of $k$ MUBs in dimension $d$ amounts to decide whether there are orthonormal bases $\aBase_1\ldots \aBase_k$ of $\mathds{C}^d$ that satisfy \eqref{eq:mubs}, where the inner product is the usual for $\mathds{C}^n$, namely
$$
\innerproduct{v}{w} = \sum_{i=1}^d v_i \overline{w_i}
$$
Considering the equations related to orthonormality conditions as well, we are left with a system of equations over $\C$ involving $\binom{kd}{2} + kd$ equations, and $kd^2$ complex variables. As explained in the previous section, we can also decompose each complex variable in its real and imaginary part and develop the equations correspondingly, obtaining a multivariate polynomial system over $\R$ involving twice as many equations and variables. Note that the system of equations over the complex variables is not a polynomial system because the conjugate operator is used in the definition of the inner product.

Taking this into account, it is then possible to compute, given $k$ and $d$, an $\Lor$-sentence $\phi_{k,d}$ stating that ``there are at least $k$ MUBs in dimension $d$''. 
The truth value of $\phi_{k,d}$ in $\R$ can be found using any available quantifier elimination  algorithm for RCF and, since $\phi_{k,d}$ only uses existential quantifiers, it can be also computed in polynomial space.%, and therefore belongs to the \textit{existential theory of the reals}, which can be decided through more efficient algorithms that exploit the particular structure of existential expressions \cite{basu2006existential}, even though in practical scenarios the general procedures for deciding the ``complete'' theory of the reals are still used \cite{passmore2009combined}. The decision problem itself is \textsc{NP-hard} and  included in PSPACE \cite{canny1988some}. 

%One of the most efficient ``implementations'' of the quantifier elimination procedure defined by Tarski is the Cylindrical Algebraic Decomposition (CAD) algorithm, developed mainly by Collins \cite{collins1975quantifier}. Even though it greatly improved the algorithm deduced by Tarski's results, it is still doubly exponential on the number of variables \cite{davenport1988real}, and therefore quite impractical for systems involving a big number of them.

In the context of the problem of determining the existence of 4 MUBs in dimension 6, the corresponding multivariate polynomial system over $\R$ has $288$ variables. Even after exploiting some symmetries (such as fixing the first base $\aBase_1$ to the canonical one and the phase of all vectors) the whole system contains $180$ variables, and therefore
is still unfeasible to solve using any of the methods described in the previous section. This motivates us to approach the problem through a heuristic algorithm.

To emphasise on the unfeasibility of employing the CAD algorithm to solve the whole formula we estimate its running time considering Renegar's bound \cite{hong1991comparison}, which states that there exists a CAD implementation whose time complexity is
$$
L(\log L)(\log \log L) (md)^{O(n)}
$$
where $L$ is the coefficients bit length (i.e.\ the number of bits required to express the coefficients of the polynomials), $m$ the number of polynomials, $d$ the total degree and $n$ the number of variables. Even assuming small hidden constants (1 for the exponent and for the whole complexity) the estimated number of operations is in the order of $10^{520}$. Note, nonetheless, that this is the worst case complexity, and considering
the particular highly symmetrical instances that arise due to the MUB conditions it might be possible to speed up the quantifier elimination procedure by some particular algorithm exploiting this structure.

\section{A heuristic algorithm to disprove the existence of $k$ MUBs in dimension $d$}\label{sec:heur}

We develop a heuristic to attempt to disprove the existence of solutions for a given polynomial system. Suppose we have a system $\{p_i(x_1,\ldots,x_k) = 0 \}_{1 \leq i \leq n}$ of $n$ polynomial equations on $k$ variables over the reals. Deciding the existence of a real root (i.e. a $\vec{z} = (z_1, \ldots, z_k) \in \mathds{R}^k$ such that $p_i(\vec{z}) = 0$ for all $1\leq i \leq n$) amounts to deciding whether the $\Lor$-sentence
\begin{equation*}
\Phi \equiv \exists x_1, \ldots, x_k \bigwedge_{i=1}^n \Phi_i   
\end{equation*}
is true in $\R$, where 
$\Phi_i \equiv p_i(x_1, \ldots, x_k) = 0$

%
%Let $\Phi_i \equiv p_i(x_1, \ldots, x_k) = 0$ be the $i$th formula and $S \subseteq \{1,\ldots,n\}$ a subset of indices for the subformulas $\Phi_i$. 
%
If a conjunction is satisifiable then any subset of the conjuncts is also satisfiable as well.
%
%if the formula $\Phi$ composed by a conjunction of subformulas is satisfiable then any subset of conjuncts of $\Phi$ is satisfiable as well)
%
Therefore, for $S \subseteq \{1,\ldots,n\}$, if $\Phi_S = \exists x_1,\ldots,x_k \bigwedge_{i\in S} \Phi_i$ is false in $\R$ we can conclude that $\Phi$ is false in $\R$ as well. Furthermore, if $|S|$ is small enough, it is possible to use the CAD algorithm to find the truth value of $\Phi_S$ in a \textit{reasonable} time.

CAD can also be used to perform quantifier elimination considering only a subset $V$ of the $k$ variables $x_1,\ldots,x_k$. Given the formula $\Phi_{S, V} = \exists \{x_j\}_{j \in V} \bigwedge_{i\in S} \Phi_i$ the algorithm will return a new formula $\Phi'$ that does not use the variables $\{x_j\}_{j \in V}$ such that $\Phi'$ is true if and only if $\Phi_{S,V}$ is true. Observe that this new formula $\Phi'$ is not necessarily of the form $q(x_1, \ldots, x_k) =0$ being $q$ a multivariate polynomial, but rather a set of equations defining the union of semialgebraic sets \cite{Basu2006}.

More formally, given a set of formulas $\Psi = \{\Phi_1, \ldots, \Phi_n\}$ over the variables $x_1, \ldots, x_k$ we define a \textit{creation step} as picking a subset $S \subseteq \{1, \ldots, n\}$ of formula indices, a subset $V \subseteq \{1, \ldots, k\}$ of variable indices and performing quantifier elimination over the sentence $\Phi_{S, V} = \exists \{x_j\}_{j \in V} \bigwedge_{i\in S} \Phi_i$ to obtain a new formula $\Phi'$. Observe that if the chosen formula $\Phi_{S, V}$ is false and has all its variables quantified then the new formula $\Phi'$ will be the \texttt{False} atom ($\Phi' = \bot$), in which case one can already conclude that the formula $\Phi = \exists x_1, \ldots, x_m \bigwedge_{i=0}^n \Phi_i$ is false as well. 

Our heuristic algorithm to disprove the existence of $k$ MUBs in dimension $d$ will define a starting set $\Psi_0$ with all formulas related to the multivariate polynomial system described in Section \ref{sec:algo}. Iteratively, we will define $\Psi_{i+1} = \Psi_i \cup \{\Phi_i'\}$ where $\Phi_i'$ is a formula obtained by applying a creation step to $\Psi_i$. $\Psi_{i+1}$ has one more formula than $\Psi_i$, and they are equal in terms of validity (the quantified conjunction of all formulas of $\Psi_i$ is true if and only if the same happens for $\Psi_{i+1}$). If for some $i$ it is the case that $\Phi_i'=\bot$ then the formula defined by the set $\Psi_i$ is false, in which case the starting $\Psi_0$ is false as well, and we conclude that there is no set of $k$ MUBs in dimension $d$. Meanwhile, if $\Phi_i' \neq \bot$, we add it to the set of formulas in order to consider it for the next creation steps. The idea is that the formulas iteratively added to $\Psi_0$ can be employed in the following creation steps to reach an ultimate contradiction by merging different simplified conditions of the original system.

If the initial set of formulas $\Psi_0$ is valid then this procedure will never halt. Meanwhile, if it is false, it might be able to find a proof of that fact represented as a deductive reduction form $\Psi_0$ to $\bot$.

We note that this scheme does not rely on the CAD algorithm, but rather on any implementation of the quantifier elimination procedure. In particular, we use the Wolfram Engine method \texttt{Resolve} that aims to reduce the given formula in any possible way (i.e.\ it might use a different quantifier elimination algorithm than CAD). Nonetheless, such method allows to force a certain strategy (for example, the algorithm defined at \cite{guangxing2004effective}).

The heuristic described above depends heavily on the way we choose the sets $S$ and $V$, specially on the size of them. We experimented to find values for them that allow \textit{1)} The \texttt{Resolve} method to finish in the order of minutes, without consuming too much memory\footnote{A big issue present in most implementations of quantifier elimination is memory consumption: in the worst case, intermediate calculations can turn out to be exponential in size.} \textit{2)} The chosen subsystem to be large and complex enough (i.e.\ involving equations that share variables) to capture relevant constraints of the original system. We observed that in order to accomplish \textit{1)} the size of $|S|$ has to depend on the number of variables present in the selected equations (as expected, considering that the CAD algorithm depends double exponentially on the number of variables), and to achieve \textit{2)} we need to quantify a \textit{reasonable} number of the variables in the chosen equations: quantifying few of them will make the resulting $\Phi'$ to be extremely long and unusable for the rest of the computations, while quantifying many of them will cause the resulting subsystem to be valid, and therefore $\Phi'$ will be the \texttt{True} atom ($\Phi' = \top$), which is useless for our purpose.

Alongside the iterations of this heuristic we will periodically perform a \textit{resolution step} (in contrast with the \textit{creation step} described above): we will take a bigger set of formulas and try to reduce them quantifying all its variables. This step is meant to take advantage of the new formulas combining them together, focusing on reaching an ultimate contradiction instead of creating new constraints.

As a proof of concept we use our algorithm to prove the nonexistence of 4 MUBs in dimension 2. In this case there are 36 equations and 16 variables. After some testing we parameterize the algorithm in the following way:

\begin{itemize}
    \item The set $S$ is picked uniformly at random, conditioning that $|S|=4$.
    \item The set $V$ is picked uniformly at random from the variables of the formulas indicated by $S$, conditioning that $|V|=6$.
    \item The \textit{resolution step} takes $13$ different formulas, and runs every 5 steps of the \textit{creation step}.
    \item All Wolfram queries have a timeout limit of 30 seconds.
\end{itemize}

On such a run, after 20 iterations we obtained a set of equations of the system that could not be satisfied. The formula can be found in the Appendix.

We conclude this section by observing that the described algorithm can be easily modified to work on weaker and easier to solve conjectures involving the existence of MUBs. For example, by removing some equations from the system one could test whether there are three MUBs in dimension 6 \textit{and} some vector $v$ that is unbiased with respect to all other vectors from the mutually unbiased bases, or even ask for independent sets of normalized vectors instead of bases (i.e.\ allowing them to contain less than $d$ vectors).

\section{Reals and other structures}\label{sec:other}
% Contar que la máxima cantidad de MUBs en los complejos basados en reales es la misma que en computables, algebraicos, definable, etc.
% https://en.wikipedia.org/wiki/Real_closed_field#Examples_of_real_closed_fields

% A field is defined as a set $F$ along with two binary operations over elements of the set, namely the addition $+$ and multiplication $\cdot$, that behave as in the real numbers:

% \begin{itemize}
%  \item Both operations are associative.
%  \item Both operations are commutative.
%  \item There exist an element of the set that is neutral under addition named $0$, and another that is neutral under multiplication called $1$.
%  \item For every $a\in F$ there exist an additive inverse $b\in F$ such that $a+b=0$.
%  \item For every $a\in F, a\neq 0$ there exist a multiplicative inverse $b$ such that $a\cdot b=1$.
%  \item Multiplication is distributable over addition.
% \end{itemize}

% We can now define a real closed field as an ordered field such that every positive element of the field has a square root in $F$, and every odd degree polynomial with coefficients in $F$ has at least one root in $F$.

A consequence of the fact mentioned in section \ref{sec:RCF} is that {\em any} $\Lor$-structure satisfying the axioms of RCF has the same truths as $\R$, and is therefore indistinguishable from $\R$ by means of $\Lor$-First-Order formulas. In the context of this work, this means that the maximum number of MUBs in dimension $d$ with coefficients with real part and imaginary part in the reals, is the same as the maximum number of MUBs in dimension $d$ with coefficients with real and imaginary part in {\em any} real closed field. Examples of Real Closed Fields other than the reals are the real algebraic numbers (those which are roots of a non-zero polynomial in one variable with integer coefficients) or the computable real numbers (those for which there is an algorithm that approximates them with a precision given as parameter to the algorithm). Since the studied properties regarding the existence of MUBs can be written in the language $\Lor$ over the reals (via the explained mapping of complex numbers to the real and imaginary part), then all our results automatically hold in any Real Closed Field, as the algebraic or computable complex numbers.

\section{Closing remarks}

In this work we presented an exponential-time algorithm based on First Order Logic tools able to decide the existence of $k$ MUBs in dimension $d$ for any values of $k,d \in \mathds{N}$. We also showed that the problem is in \textsc{PSPACE}, since it reduces to deciding the truth value of a formula from the theory of the Real Closed Fields using only existential quantification (the ``Existential theory of the reals'').

Since this algorithm requires an enormous amount of time to solve the decision problem even when $d=6$ we defined an heuristic approach to design a semi-decision procedure that can detect whether there are not $k$ mutually unbiased bases in dimension $d$, and that will not halt if they exist. This algorithm does not actually exploit any particularities of the MUB problem, rather than the fact that it can be expressed in the RCF logic. We implemented the defined heuristic and provided a proof of concept by using it to show that there are not $4$ MUBs in dimension $2$.

As a byproduct of these results it can be proved that, given any model $M$ of RCF (such as $\mathds{R}$ or the algebraic reals) if $m$ is the maximum number of MUBs in any dimension $d$, then there is a set $\{\aBase_1, \ldots, \aBase_m\}$ of mutually unbiased bases such that the real and imaginary part of all imaginary numbers involved in the vectors from $\aBase_1, \ldots, \aBase_m$ all belong to $M$. More formally, for any $1\leq i \leq m$ and $v \in \aBase_i$ it is the case that $\Re (v_j), \Im (v_j) \in M$ for $1 \leq j \leq d$. This implies that when studying the maximum number of MUBs in any dimension there is no loss of generality in assuming that the coefficients of the vectors belong to any simpler model of RCF, limiting therefore their algebraic complexity.

Finally, we conclude this work by noting that the decidability of other quantum information problems has already been addressed, obtaining both positive and negative results \cite{cubitt2015undecidability, wolf2011problems}.

\newpage
\section{Appendix}

In our implementation the bases $\{\aBase_1, \aBase_2, \aBase_3, \aBase_4\}$ are denoted with variables $x,y,z$ and $w$, and subindices are used to refer to the particular components of the vectors. For example, the real part of the $i$th component of the $k$th vector from the base $y$ is represented by the variable $yki0$, while the complex part is represented by $yki1$. The base $\aBase_1$ is fixed to be the canonical one, and it is assumed that every vector is phase shifted in such a way that the first component is a real number, and therefore the corresponding imaginary part is 0.

During the \textit{creation steps} one of the formulas that was created is the following one:

\begin{lstlisting}[caption=Example of a formula deduced by performing quantifier elimination over a subset of formulas and variables., captionpos=b, label=lst:creation_step]
    (w011 < 0 AND -(1/Sqrt[2]) <= w111 <= 1/Sqrt[2] AND w110 == -(Sqrt[1 - 2*w111^2]/Sqrt[2])) OR (w110 == Sqrt[1 - 2*w111^2]/Sqrt[2]) OR (w011 == 0 AND w010 < 0 AND w111 == -(1/Sqrt[2]) && w110 == 0) OR ... 
\end{lstlisting}

This expression represents, as explained in Section \ref{sec:heur}, the union of some semialgebraic sets.

The subset of formulas that was found to forbid the existence of 4 MUBs in dimension 2 included the previous one, as well as some others such as

\begin{lstlisting}[caption=Examples of initial equations that were used in the last resolution step to reach an ultimate contradiction, captionpos=b, label=lst:resol_step]
    (1/Sqrt[2]*1/Sqrt[2]+w010*w110+w011*w111)^2+(w011*w110-w010*w111)^2==0

    (1/Sqrt[2]*1/Sqrt[2]+z110*z110+z111*z111)^2+(z111*z110-z110*z111)^2==1

    (1/Sqrt[2]*1/Sqrt[2]+y110*z110+y111*z111)^2+(y111*z110-y110*z111)^2==1/2
\end{lstlisting}

The formulas listed in \ref{lst:resol_step} correspond, in order, with the orthogonality conditions between vectors of the same base, the normality condition, and the unbiased equation relating vectors of different bases.

Note that these equations are already simplified using the symmetries described in Section \ref{sec:algo}. For instance, in the orthogonality condition it can be appreciated that \texttt{w000} and \texttt{w100} are assumed to be $\frac{1}{\sqrt{2}}$, while \texttt{w001} and \texttt{w101} are 0.

\bibliographystyle{plain}
\bibliography{biblio}

\begin{thebibliography}{10}

\bibitem{Adamson2010}
RBA Adamson and Aephraim~M Steinberg.
\newblock Improving quantum state estimation with mutually unbiased bases.
\newblock {\em Physical review letters}, 105(3):030406, 2010.

\bibitem{Archer2005}
Claude Archer.
\newblock There is no generalization of known formulas for mutually unbiased bases.
\newblock {\em Journal of mathematical physics}, 46(2):022106, 2005.

\bibitem{Bae2022}
Joonwoo Bae, Anindita Bera, Dariusz Chru{\'s}ci{\'n}ski, Beatrix~C Hiesmayr, and Daniel McNulty.
\newblock How many mutually unbiased bases are needed to detect bound entangled states?
\newblock {\em Journal of Physics A: Mathematical and Theoretical}, 55(50):505303, 2022.

\bibitem{Bae2019}
Joonwoo Bae, Beatrix~C Hiesmayr, and Daniel McNulty.
\newblock Linking entanglement detection and state tomography via quantum 2-designs.
\newblock {\em New Journal of Physics}, 21(1):013012, 2019.

\bibitem{Bandyopadhyay2002}
Somshubhro Bandyopadhyay, P~Oscar Boykin, Vwani Roychowdhury, and Farrokh Vatan.
\newblock A new proof for the existence of mutually unbiased bases.
\newblock {\em Algorithmica}, 34(4):512--528, 2002.

\bibitem{Basu2006}
Saugata Basu, Richard Pollack, and Marie-Françoise Roy.
\newblock {\em Algorithms in Real Algebraic Geometry}.
\newblock Springer, 2006.

\bibitem{Bengtsson2007}
Ingemar Bengtsson.
\newblock Three ways to look at mutually unbiased bases.
\newblock In {\em AIP Conference Proceedings}, volume 889, pages 40--51. American Institute of Physics, 2007.

\bibitem{Bengtsson2007b}
Ingemar Bengtsson, Wojciech Bruzda, {\AA}sa Ericsson, Jan-{\AA}ke Larsson, Wojciech Tadej, and Karol {\.Z}yczkowski.
\newblock Mutually unbiased bases and hadamard matrices of order six.
\newblock {\em Journal of mathematical physics}, 48(5):052106, 2007.

\bibitem{Brierley2008}
Stephen Brierley and Stefan Weigert.
\newblock Maximal sets of mutually unbiased quantum states in dimension 6.
\newblock {\em Physical Review A}, 78(4):042312, 2008.

\bibitem{Brierley2009}
Stephen Brierley and Stefan Weigert.
\newblock Constructing mutually unbiased bases in dimension six.
\newblock {\em Physical Review A}, 79(5):052316, 2009.

\bibitem{Brierley2010}
Stephen Brierley and Stefan Weigert.
\newblock Mutually unbiased bases and semi-definite programming.
\newblock In {\em Journal of Physics: Conference Series}, volume 254, page 012008. IOP Publishing, 2010.

\bibitem{Butterley2007}
Paul Butterley and William Hall.
\newblock Numerical evidence for the maximum number of mutually unbiased bases in dimension six.
\newblock {\em Physics Letters A}, 369(1-2):5--8, 2007.

\bibitem{Coles2017}
Patrick~J Coles, Mario Berta, Marco Tomamichel, and Stephanie Wehner.
\newblock Entropic uncertainty relations and their applications.
\newblock {\em Reviews of Modern Physics}, 89(1):015002, 2017.

\bibitem{collins1975quantifier}
George~E Collins.
\newblock Quantifier elimination for real closed fields by cylindrical algebraic decompostion.
\newblock In {\em Automata Theory and Formal Languages: 2nd GI Conference Kaiserslautern, May 20--23, 1975}, pages 134--183. Springer, 1975.

\bibitem{cubitt2015undecidability}
Toby~S Cubitt, David Perez-Garcia, and Michael~M Wolf.
\newblock Undecidability of the spectral gap.
\newblock {\em Nature}, 528(7581):207--211, 2015.

\bibitem{Durt2010}
Thomas Durt, Berthold-Georg Englert, Ingemar Bengtsson, and Karol {\.Z}yczkowski.
\newblock On mutually unbiased bases.
\newblock {\em International journal of quantum information}, 8(04):535--640, 2010.

\bibitem{Friis2019}
Nicolai Friis, Giuseppe Vitagliano, Mehul Malik, and Marcus Huber.
\newblock Entanglement certification from theory to experiment.
\newblock {\em Nature Reviews Physics}, 1(1):72--87, 2019.

\bibitem{guangxing2004effective}
Zeng Guangxing and Zeng Xiaoning.
\newblock An effective decision method for semidefinite polynomials.
\newblock {\em Journal of Symbolic Computation}, 37(1):83--99, 2004.

\bibitem{hong1991comparison}
Hoon Hong et~al.
\newblock {\em Comparison of several decision algorithms for the existential theory of the reals}.
\newblock Citeseer, 1991.

\bibitem{Huang2021}
Chang-Jiang Huang, Guo-Yong Xiang, Yu~Guo, Kang-Da Wu, Bi-Heng Liu, Chuan-Feng Li, Guang-Can Guo, and Armin Tavakoli.
\newblock Nonlocality, steering, and quantum state tomography in a single experiment.
\newblock {\em Physical Review Letters}, 127(2):020401, 2021.

\bibitem{Ivonovic1981}
ID~Ivonovic.
\newblock Geometrical description of quantal state determination.
\newblock {\em Journal of Physics A: Mathematical and General}, 14(12):3241, 1981.

\bibitem{Jaming2009}
Philippe Jaming, M{\'a}t{\'e} Matolcsi, P{\'e}ter M{\'o}ra, Ferenc Sz{\"o}ll{\H{o}}si, and Mih{\'a}ly Weiner.
\newblock A generalized pauli problem and an infinite family of mub-triplets in dimension 6.
\newblock {\em Journal of Physics A: Mathematical and Theoretical}, 42(24):245305, 2009.

\bibitem{Kibler2006}
Maurice~R Kibler and Michel Planat.
\newblock A su (2) recipe for mutually unbiased bases.
\newblock {\em International Journal of Modern Physics B}, 20(11n13):1802--1807, 2006.

\bibitem{Klappenecker2004}
A~Klappenecker and M~R{\"o}tteler.
\newblock Finite fields and applications.
\newblock {\em Lecture Notes in Computer Science}, 2948:137--144, 2004.

\bibitem{marker2006model}
David Marker.
\newblock {\em Model theory: an introduction}, volume 217.
\newblock Springer Science \& Business Media, 2006.

\bibitem{papadimitriou2003computational}
Christos~H Papadimitriou.
\newblock Computational complexity.
\newblock In {\em Encyclopedia of computer science}, pages 260--265. 2003.

\bibitem{Paterek2009}
Tomasz Paterek, Borivoje Daki{\'c}, and {\v{C}}aslav Brukner.
\newblock Mutually unbiased bases, orthogonal latin squares, and hidden-variable models.
\newblock {\em Physical Review A}, 79(1):012109, 2009.

\bibitem{Paul2016}
EC~Paul, DS~Tasca, {\L}ukasz Rudnicki, and SP~Walborn.
\newblock Detecting entanglement of continuous variables with three mutually unbiased bases.
\newblock {\em Physical Review A}, 94(1):012303, 2016.

\bibitem{Raynal2011}
Philippe Raynal, Xin L{\"u}, and Berthold-Georg Englert.
\newblock Mutually unbiased bases in six dimensions: The four most distant bases.
\newblock {\em Physical Review A}, 83(6):062303, 2011.

\bibitem{Seyfarth2011}
Ulrich Seyfarth and Kedar~S Ranade.
\newblock Construction of mutually unbiased bases with cyclic symmetry for qubit systems.
\newblock {\em Physical Review A}, 84(4):042327, 2011.

\bibitem{Spengler2012}
Christoph Spengler, Marcus Huber, Stephen Brierley, Theodor Adaktylos, and Beatrix~C Hiesmayr.
\newblock Entanglement detection via mutually unbiased bases.
\newblock {\em Physical Review A}, 86(2):022311, 2012.

\bibitem{Wehner2010}
Stephanie Wehner and Andreas Winter.
\newblock Entropic uncertainty relations—a survey.
\newblock {\em New Journal of Physics}, 12(2):025009, 2010.

\bibitem{Wiesniak2011}
M~Wie{\'s}niak, Tomasz Paterek, and Anton Zeilinger.
\newblock Entanglement in mutually unbiased bases.
\newblock {\em New Journal of Physics}, 13(5):053047, 2011.

\bibitem{wolf2011problems}
Michael~M Wolf, Toby~S Cubitt, and David Perez-Garcia.
\newblock Are problems in quantum information theory (un) decidable?
\newblock {\em arXiv preprint arXiv:1111.5425}, 2011.

\bibitem{Wootters1989}
William~K Wootters and Brian~D Fields.
\newblock Optimal state-determination by mutually unbiased measurements.
\newblock {\em Annals of Physics}, 191(2):363--381, 1989.

\bibitem{Wu2009}
Shengjun Wu, Sixia Yu, Klaus M{\o}lmer, et~al.
\newblock Entropic uncertainty relation for mutually unbiased bases.
\newblock {\em Physical Review A}, 79(2):022104, 2009.

\bibitem{Yuan2016}
Hao Yuan, Zheng-Wei Zhou, and Guang-Can Guo.
\newblock Quantum state tomography via mutually unbiased measurements in driven cavity qed systems.
\newblock {\em New Journal of Physics}, 18(4):043013, 2016.

\end{thebibliography}

% \onecolumn\newpage
\end{document}